# Automatisierte Verwaltung von ITS Roadside Stations für den sim$^{TD}$ Feldversuch

**Autoren**
Prof. Dr. Horst Wieker, Bechir Allani, Thomas Baum, Manuel Fünfrocken, Arno Hinsberger, Jonas Vogt, Sebastian Weber

**Organisation**
Hochschule für Technik und Wirtschaft des Saarlandes,
Forschungsgruppe Verkehrstelematik
Goebenstraße 40, 66117 Saarbrücken

**Themenfeld**: Automatisierungssysteme/Assistenzsysteme -> Kommunikation, C2X
**Schlagwörter**: IRS, C2I, sim$^{TD}$, AKTIV
**Neuigkeitsgrad**: Weltneuheit

## Abstract

Das Projekt sim$^{TD1}$ ist der erste große Feldversuch für Fahrzeug-zu-Fahrzeug und Fahrzeug-zu-Infrastruktur Kommunikation in Europa. Es besteht aus bis zu 400 Fahrzeugen und über 100 infrastrukturseitigen Kommunikationseinheiten, sog. ITS Roadside Stations (IRS). Bei der großen Anzahl von abgesetzten Einheiten wird ein leistungsfähiges Verwaltungssystem benötigt, das sicherstellt, dass alle notwendigen Verwaltungsaufgaben für die IRS ausgeführt werden können: von der grundlegenden Konfiguration, über die Installation und Verwaltung von Anwendungen bis hin zur Behandlung und Behebung von Fehlern der IRS selbst. Des Weiteren soll eine grafische Oberfläche für die Administration geschaffen, ein verschlüsselter Kommunikationskanal implementiert und ein Framework für die Anwendungen von Drittanbietern entwickelt werden. Aufgrund der Wichtigkeit des Managements für das gesamte Projekt muss das Managementsystem hoch verfügbar sein.

## Einführung

Bisher haben europäische Forschungsprojekte lediglich kleinere Integrationsprojekte und Machbarkeitsstudien durchgeführt. Das Ziel dieser Projekte[2] war nie, zu zeigen, dass die V2X[3] Kommunikation einen signifikanten Einfluss auf die Verkehrseffizienz und die Verkehrssicherheit hat; dies wurde bisher nur mit Hilfe von Simulationen oder Potenzialanalysen nachgewiesen. Im Projekt sim$^{TD}$, welches staatlich gefördert wird, ist die Evaluierung der verkehrlichen Wirkung ein Hauptziel. In sim$^{TD}$ haben sich deutsche

---

[1] Sichere Intelligente Mobilität – Testfeld Deutschland
[2] Beispielsweise in folgenden Projekten: [2], [10], [11], [12], [13], [14], [15]
[3] V2X bezeichnet sowohl die Fahrzeug-zu-Fahrzeug (engl.: Vehicle-to-Vehicle) als auch die Fahrzeug-zu-Infrastruktur (engl. Vehicle-to-Infrastructur) Kommunikation

Automobilhersteller, Zulieferer, Netzprovider, öffentliche Einrichtungen und führende Forschungseinrichtung zusammengeschlossen, um dieses Ziel zu verwirklichen. Eines der Hauptziele ist, die Möglichkeiten und Vorzüge von V2X Anwendungen im Zusammenspiel mit den realen, verkehrlichen Wirksystemen zu zeigen. Ausführliche Versuche und Tests werden auf Autobahnen und Landstraßen in Hessen und in einem ausgewählten Stadtteil in Frankfurt am Main stattfinden. Eine Übersicht des Versuchsgebietes ist in Abbildung 1 dargestellt. Es ist in vier Bereiche untergliedert, wobei jeder dieser Bereiche für eine spezifische Umgebung mit ihren spezifischen Eigenschaften steht. Im Einzelnen sind dies: Autobahnen mit hoher IRS Dichte (dunkelblau), Autobahnen mit geringer IRS Dichte (blau), Landstraßen (grün) und ein innerstädtisches Gebiet (rot).

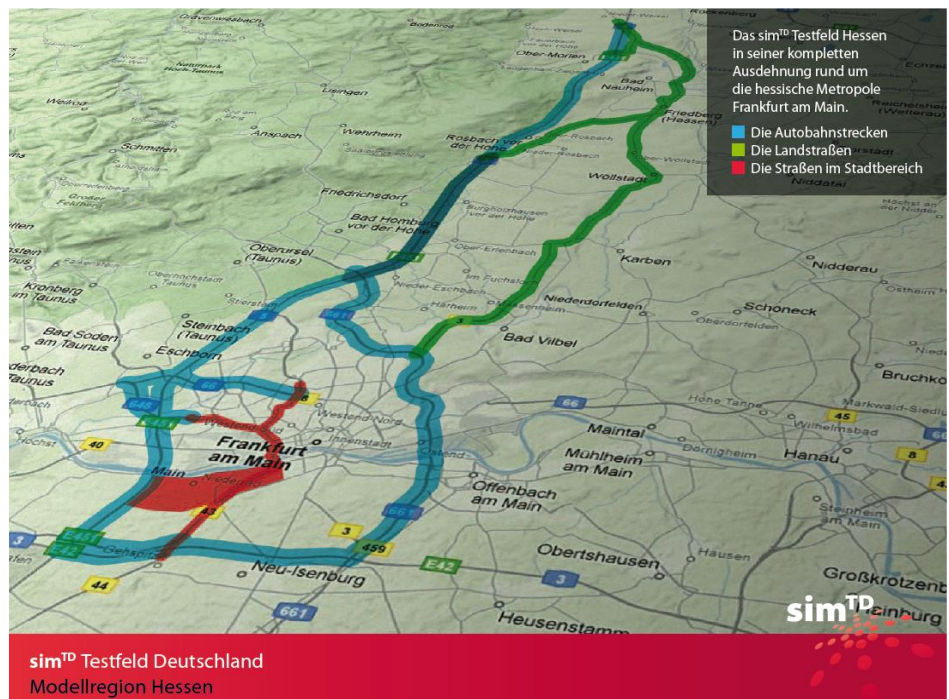

**Abbildung 1: sim$^{TD}$ Versuchsgebiet [17]**

Den IRS kommt dabei die Aufgabe zu, Fahrzeuge sowohl mit den klassischen als auch mit der zukünftigen Infrastruktur – inklusive der Verkehrszentralen – zu verbinden. Sie spielen in diesem Szenario generell, aber vor allem während der Einführungsphase, eine entscheidende Rolle.

## IRS Management System

In einer V2I Umgebung sind die IRS für die Verbindung zwischen den Fahrzeugen, der am Straßenrand stehenden Infrastruktur und den Verkehrs(-management)zentralen verantwortlich. Darüber hinaus sind sie in der Lage den Sendebereich von Fahrzeugen zu erweitern und tragen dadurch dafür Sorge, dass relevante Nachrichten länger in ihrem Relevanzgebiet verfügbar sind. Dies geschieht dabei unabhängig von der Dichte des Funknetzwerkes durch Techniken wie z.B. „Store and Forward". Durch die hohe Anzahl von Nachrichten und die sich daraus ergebenden Netzwerk- und Ressourcennutzung ist eine dezentralisierte Vorverarbeitung und Aggregation von Nachrichten unabdingbar. Zur Sicherstellung der Skalierbarkeit ist es notwendig, dass infrastrukturseitige Anwendungen[4] eine verteilte hierarchische Struktur besitzen. Dafür stellt die IRS eine Plattform für beliebige

---
[4] Anwendungen werden im sim$^{TD}$ Jargon Funktionen genannt.

Arten von V2I Anwendungen zur Verfügung. Diesen Anwendungen stehen für ihre Arbeit abstrahierte Kommunikationsschnittstellen und verschiedene Systemressourcen zur Verfügung.

Um die Verfügbarkeit und die Funktionalität der IRS zu garantieren, ist ein Managementsystem für IRS unabdingbar. Da ein schneller und einfacher physikalischer Zugang zu den IRS nicht möglich ist, ist es notwendig, dass alle Managementfunktionalitäten remote (d.h. nicht vor Ort) durchgeführt werden können. Um die Komplexität des Managementsystems so gering wie möglich zu halten und gleichzeitig die Verfügbarkeit zu erhöhen, wurde es in verschiedene, interagierende Subsysteme und Module unterteilt. Jedes dieser Systeme hat dabei seine definierte und dedizierte Aufgabe:

- Das *Configuration Management Subsystem* ist für die Installation von Anwendungen und des Systems selbst verantwortlich.
- Das *Fault Management Subsystem* ist verantwortlich für die qualifizierte Behandlung von Ausnahmesituationen und Fehlern.
- Das *Function Framework* bietet den Anwendungen einen definierten und kontrollierten Zugriff auf Schnittstellen (z.B. Kommunikationsschnittstellen oder Systemressourcen).

## IRS Management Center Architektur

Um die obigen Ziele zu gewährleisten, ist es notwendig, ein skalierbares und hochverfügbares IRS Management Center aufzubauen.
In Abbildung 2 ist die Hardwarearchitektur dieses Management Centers dargestellt. Die redundante und mit Load Balancer ausgestattete Lösung sorgt in Verbindung mit der eingesetzten Virtualisierung für eine robuste Architektur und reduziert gleichzeitig die Recovery Time nach einem Hardwarefehler.

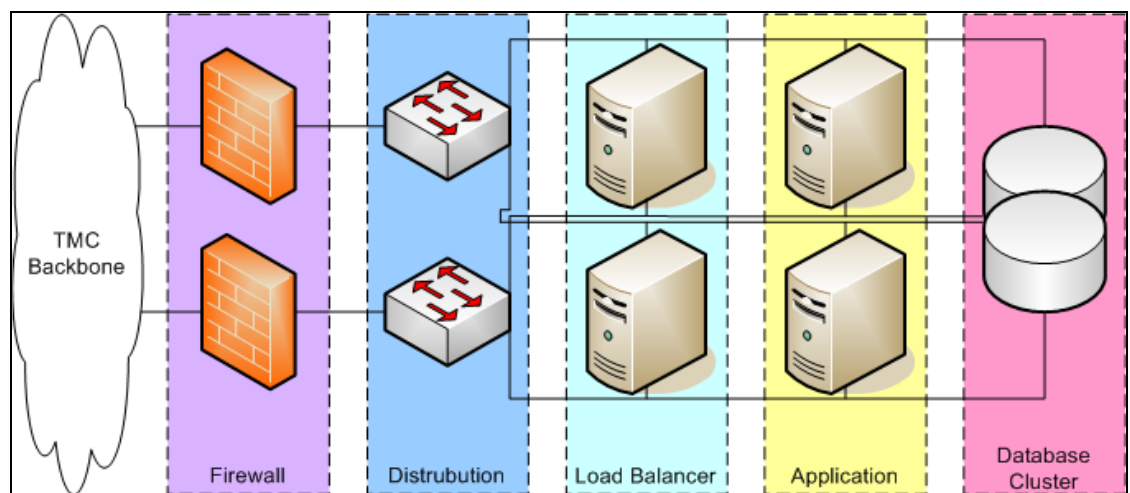

**Abbildung 2: IRS Management Center (physikalisch)**

Der gesamte eingehende Netzwerkverkehr im Management Center wird durch eine Firewall gescannt und über die Verteilungsebene (Distribution) an die Load Balancer weitergeleitet. Die Verbindungen werden dann von den Load Balancer gleichmäßig auf die Managementserver (Application) verteilt. Die Management-Server sind alle mit der gleichen Hard- und Software ausgestattet. Die Verarbeitungsrechenleistung kann mit diesem Konzept durch das Hinzufügen eines weiteren Servers einfach erweitert werden. Dies – in Zusammenarbeit mit dem Datenbankcluster – führt zu einem hohen Grad an Skalierbarkeit. Im Datenbankcluster sind die Konfigurations- und Managementdaten für die Management-

Server abgelegt. Das Gesamtkonzept stellt damit sicher, dass der Ausfall einer Hardwarekomponente nicht zu einem kritischen Datenverlust oder einer Nichtverfügbarkeit des Management Centers führen kann. Darüber hinaus ist das System derart konzipiert, dass es bei einer steigenden Zahl von IRS entsprechend skaliert.

## IRS Anwendungsplattform

Die IRS Anwendungsplattform ist in vier Subsysteme (siehe Abbildung 4) unterteilt:
- *Configuration Management*
- *Fault Management*
- *Function Framework*
- Kommunikationsmodule

Die Kommunikationsmodule stellen dabei folgende Schnittstellen zur Verfügung:
- V2I Kommunikation, die für die Kommunikation mit den Fahrzeugen verantwortlich ist.
- Versuchszentralenkommunikation, die Kommunikation sowohl zu zentralenseitigen Funktionsanteilen als auch zum Managementsystem zur Verfügung stellt.
- Infrastrukturkommunikation, diese ermöglicht eine Kommunikation mit z.B. Lichtsignalanlagen (LSA) oder Verkehrszeichenbrücken.

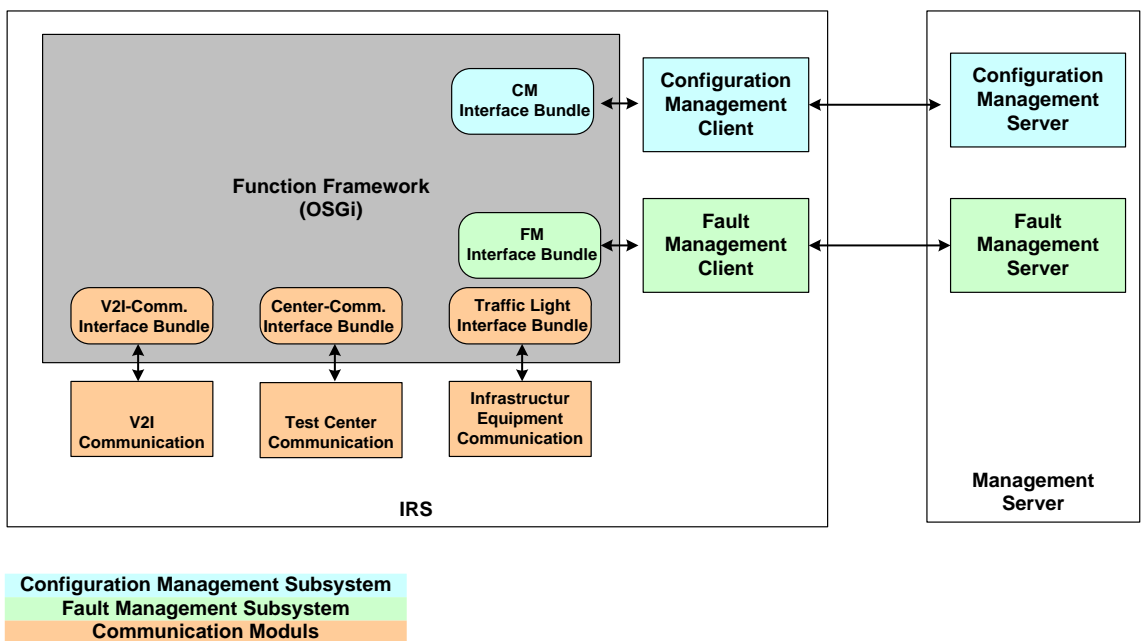

**Abbildung 3: IRS Anwendungsplattform**

## Management Subsysteme

In Folgenden werden die oben genannten vier Module: *Configuration Management*, *Fault Management*, *Function Framework* und die Kommunikationsmodule detailliert beschrieben.

### Configuration Management
Im Projekt sim$^{TD}$ ist es möglich, dass nicht alle Anwendungen auf allen IRS installiert bzw. gleichzeitig aktiv sind. So sind z.B. Anwendungen mit LSA Kommunikation auf Autobahnen zumeist nicht von Bedeutung. Daraus ergibt sich die Notwendigkeit einzelne IRS dediziert ansprechen und konfigurieren zu können. Insbesondere wird hierdurch die (Re)Installation, die (Re)Konfiguration und die (Re)Aktivierung von Anwendungen – abhängig von den jeweiligen Versuchsszenarien – ermöglicht.

Darüber hinaus muss das *Configuration Management* in der Lage sein, sowohl die Systemkonfiguration als auch die Konfiguration von Funktionen anpassen zu können. Da die IRS als flexible Anwendungsplattform für jedwede Art von V2I Anwendung entworfen ist, muss das CM Funktionen und Systemfunktionalitäten auf dieselbe Art und Weise behandeln können. Dieser Ansatz erlaubt eine effiziente und flexible Implementierung eines Managementsystems. Zur Erreichung dieser Ziele nutzt das CM ein intelligentes Paketmanagementsystem und Konfigurationsframework.

Um die zuvor genannten Ziele zu erreichen ist das CM aus zwei koexistierenden Subsystemen aufgebaut:
- *Application and Resources*: Verantwortlich für die Behandlung und Verarbeitung von Anwendungen und deren Ressourcen wie beispielsweise Bibliotheken oder Bildern.
- *Configuration*: Verteilt und verwaltet die *Properties* und sonstigen Konfigurationen für alle Anwendungen auf allen IRS.

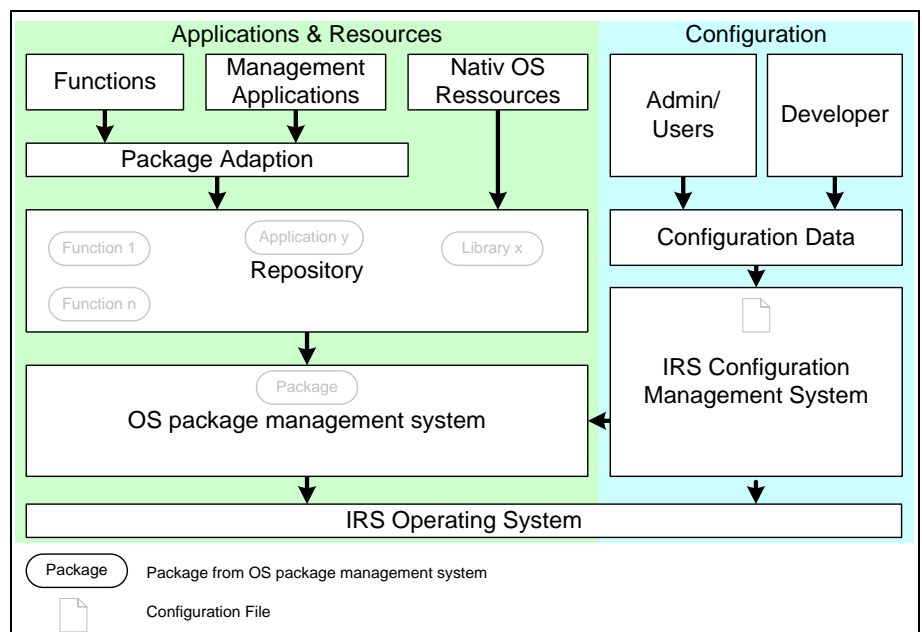

**Abbildung 4: Configuration Management**

Wie man in Abbildung 4 erkennen kann, werden – für jeden Teil getrennt – Eingaben aus verschiedenen Quellen auf ein einheitliches Format gebracht. Das bedeutet, dass Daten aus unterschiedlichen Quellen in ein einheitliches Containerformat gebracht und dabei standardisiert und aggregiert werden. Im Bereich *Application and Resources* bedeutet dies die Erstellung eines Paketes, das konform zum Paketverwaltungssystem des darunterliegenden Betriebssystems ist. Diese Aggregation erlaubt es mit dem Betriebssystempaketverwaltungssystem sowohl native Betriebssystempakete als auch sim$^{TD}$ spezifische Anwendungs- bzw. Managementpakete zu verwalten. Der gewählte Ansatz verringert nicht nur die Komplexität des Managementsystems, sondern erhöht auch Gleichzeitig die Wart- und Erweiterbarkeit des Systems.

Die Konfigurationsdaten für alle Anwendungen und die IRS werden im zentralen Datenbankcluster abgelegt. Dies erlaubt eine Entkopplung des Konfigurationsprozesses von den Prozessen zur Eingabe der Konfigurationsdaten. Daraus folgt, dass es für eine

(Re)Konfiguration unerheblich ist, ob die Daten durch einen menschlichen Administrator/Operator, durch ein anderes Subsystem oder durch eine sonstige Quelle eingefügt bzw. geändert wurden. Darüber hinaus erlaubt dies die persistente Speicherung der Konfigurationsdaten einer IRS sogar für den Fall, dass eine IRS ausgetauscht werden muss, z.B. bei einer Mobilen IRS im Zusammenhang mit einem Baustellenszenario. Im Sinne der Konfiguration und Administration bedeutet dies, dass es möglich ist eine ausgetauschte IRS automatisch zu reinstallieren und zu rekonfigurieren.

**Fault Management**

Das *Fault Management* ist, wie das *Configuration Management,* in verschiedene Subsysteme unterteilt. Es besteht daher ebenfalls aus interaktiven IRS- und zentralseitigen Teilsystemen.

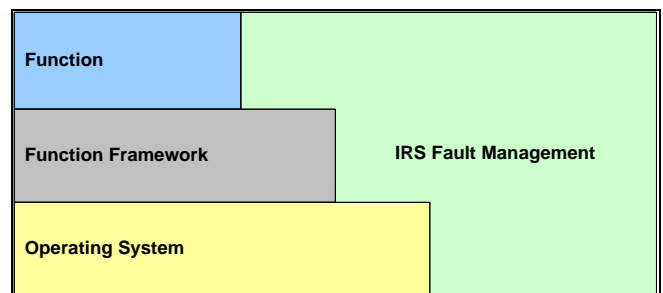

**Abbildung 5: IRS-seitiges Fault Managements**

Auf der IRS ist das Fault Management schon während der Startphase des Betriebssystems aktiv und überwacht darüber hinaus die Integration des *Function Frameworks* und der Anwendungen. Abbildung 5 zeigt diese drei Teilsysteme für die das Fault Management verantwortlich ist.

Das Fault Management hat in sim$^{TD}$ folgende Aufgaben:
- Klassifizierung und Definition von Fehlern
- Lokale Systemverifikation
- Netzwerkmanagement
- Analyse der Logdaten
- Test der Systemkomponenten
- Überwachung der Datensammlung
- Fehlerbeseitigung, durch Anwendung von vordefinierten Strategien

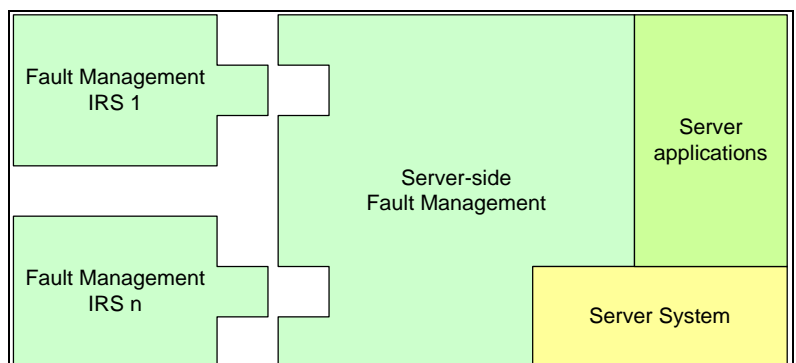

**Abbildung 6: Zentralseitiges Fault Management**

Der zentralseitige Anteil des *Fault Managements* ist als Gegenstück zum IRS-seitigen *Fault Management* aufgebaut. Die Verantwortlichkeiten sind in zwei Teilsysteme unterteilt: Auf der einen Seite die Überwachung der serverseitigen Prozesse und auf der anderen Seite die

Zusammenfassung und Aufbereitung von Nachrichten des IRS Fault Managements. Das zentralseitige *Fault Management* ist somit ein Dienstanbieter und eine Entscheidungsinstanz für das IRS-seitige Fault Management.

Alle Tätigkeiten des gesamten Managementsystems und seiner Prozesse werden durch das *Fault Management* überwacht. Zusätzlich werden alle Systeminformationen zentral gespeichert und stehen somit für eine nachträgliche und ausführliche Systemanalyse zur Verfügung.

**Function Framework**
Die Funktionen, die auf der IRS Plattform ausgeführt werden, benutzen verschiedene Schnittstellen und verbrauchen unterschiedlich viele Systemressourcen. Das IRS Management stellt zu diesem Zweck sicher, dass alle Funktionen die Schnittstellen und Systemressourcen (CPU-Zeit, RAM, Festplattenplatz, Bandbreite, etc.) gleichberechtigt nutzen können, insbesondere ohne, dass diese überlastet werden.

Für einen zuverlässigen Betrieb ist es essenziell, dass die (abgesetzten) IRS Einheiten zu jeder Zeit vom zentralen Management aus erreichbar sind. Das bedeutet es muss garantiert werden, dass ein definierter Teil der Systemressourcen und der Bandbreite für Managementaufgaben bzw. -kommunikation zur Verfügung steht. IRS Funktionen selbst haben unterschiedliche Prioritäten. Daher muss das Management garantieren, dass hochpriorisierte Funktionen gegenüber niedrig priorisierten Funktionen bei der Nutzung von Systemressourcen bevorzugt behandelt werden. Um dies garantieren zu können werden Funktionen Limits bzgl. ihrer nutzbaren Bandbreiten und Systemressourcen zugeteilt. Das *Function Framework* ist in diesem Prozess unverzichtbar für das Management um definierte, kontrollierbare und konsistente Schnittstellen für jedwede Art von Ressourcen für Anwendungen zur Verfügung stellen zu können. Ein zweites, zusätzliches und unabhängiges *Framework* beinhaltet die Managementkomponenten. Dies garantiert, dass das Management funktionsfähig bleibt, auch wenn das *Function Framework* fehlerhaft ist.

Das *Function Framework* ist als OSGi [9] Framework realisiert, wobei jede durch ein sogenanntes Bundle repräsentiert wird. Da es bei dem OSGi Framework um eine Service-orientierte Architektur handelt, ist jede Schnittstelle als Dienst realisiert. Die Funktionen sind Konsumenten der von den Schnittstellen zur Verfügung gestellten Dienste und umgekehrt.

**Kommunikation**
Bei einer IRS ist die Bandbreite auf allen Kommunikationsverbindungen beschränkt; dies betrifft insbesondere die Funkschnittstelle zu den Fahrzeugen aber auch die Verbindung zur Zentrale. Aus diesem Grund ist die jeder Anwendung zur Verfügung gestellte Bandbreite auf der IRS beschränkt. Die dafür notwendige Konfiguration für jede einzelne Anwendung ist in der zentralen Datenbank im IRS Management Center abgelegt. Die Änderungen daran dürfen nur von Managementinstanzen durchgeführt werden. Die Konfigurationen selbst finden dann auf der IRS Anwendung. Dieser Mechanismus stellt sicher, dass die Bandbreite einer bestimmten Verbindung nicht durch eine einzelne Anwendung belegt werden kann.

Die zur Verfügung stehende Bandbreite variiert von IRS zu IRS und ist abhängig vom Kommunikationsmedium, der Anzahl der Anwendungen auf einer IRS und den genutzten Protokollen. Die Verbindungen zur Zentrale können dabei privat oder öffentlich, exklusive oder geteilte physikalische Verbindungen sein. Die IRS muss also in der Lage sein sich adaptiv auf die Bedingungen der Verbindung, wie etwa Bandbreite oder Delay, einzustellen. In sim$^{TD}$ kommen dafür voraussichtlich drei Arten von Zugangstechnologien zum Einsatz:

Lichtwellenleiter (LWL), Universal Mobile Telecommunications System (UMTS) bzw. General Packet Radio Service (GPRS) und Digital Subscriber Line (xDSL).

Die aufgeführten Kommunikationsmedien haben jeweils sehr unterschiedliche Bandbreiten[5] und ebenso unterschiedliche Delayzeiten. Die benötigten Limitierungen für die Kommunikation werden dabei immer auf die tatsächlich genutzte Kommunikationstechnologie angepasst. Der Hauptgrund diese Bandbreitenbeschränkung ist die Sicherstellung der Wartbarkeit der IRS, da keine Anwendungen in der Lage ist die gesamte Bandbreite der Verbindung zur Zentrale zu belegen; und damit eine Verbindung von einer IRS zum zentralen IRS Management Center jederzeit möglich ist.

Die IRS stellt ein wichtiges Bindeglied bei einer Kommunikation zwischen zentralseitigen und fahrzeugseitigen Funktionsanteilen zur Verfügung. Dafür müssen die verschiedenen Kommunikationssysteme überbrückt werden, welche vor allem wegen Verkehrssicherheits- und Verkehrseffizienzanwendungen [3] entkoppelt sein müssen. Die zentralseitige Infrastrukturkommunikation basiert auf verbindungsorientierter TCP/IP Punkt-zu-Punkt Kommunikation wohingegen die V2I Kommunikation ihren Informationsaustausch an die aktuell verfügbaren Nachbarn und die geringe Bandbreite anpassen muss [4].

Um die Kompatibilität zu gewährleisten basiert das aufgezeigte IRS System auf der *ITS Station Reference Architecture* wie in [15] und [16] definiert. Dies setzt voraus, dass die Standard Kommunikationsmechanismen wie z.B. das Network Layer Routing und der Transport von Nachrichten ebenso wie die *Application Support Facilities* (CAM, DENM, etc.) auf beiden Plattformen IRS und IVS[6] identisch sind. Der Nachrichtentransport von der Zentrale an eine bestimmte Gruppe von Fahrzeugen erfordert auf der IRS architekturerweiternde, zusätzliche und intelligente Puffer- und Verteilungsalgorithmen. Eine intelligente Verteilung von Informationen aus einer Zentrale in ein Fahrzeugnetzwerk bedarf der Anpassung an die aktuelle Verkehrslage, die lokalen Kanaleigenschaften, die Anwendungsprioritäten und die Nachrichteneigenschaften und sichert gleichzeitig eine definiertes Redundanzniveau.

In sim$^{TD}$ kommen zwei Wireless-Technologien zum Einsatz: IEEE 802.11p (pre-standard) [7] für Verkehrssicherheits- und Verkehrseffizienzanwendungen basierend auf V2V und V2I Kommunikation und IEEE 802.11b/g [6] Wireless LAN für serviceorientierte Kommunikation in einem V2I Szenario sowie das Übertragen von Mess- bzw. Logdaten, die für die sim$^{TD}$ Versuchsauswertung herangezogen werden.

Alle diese Kommunikationstechnologien haben etwas gemeinsam: sie werden benutzt, um sensitive Informationen zu verteilen, welche authentifiziert oder besser noch verschlüsselt werden sollten. Die verschiedenen Kommunikationsarten fordern verschiedene Sicherheitsstufen und -mechanismen. Für die IEEE 802.11p Kommunikation werden Sicherheitsmechanismen benutzt, die auf einer für sim$^{TD}$ adaptierten Variante des IEEE pre-standard 1609.2 [5] beruhen [18]. Es ist geplant diese Technik ebenfalls für die IEEE 802.11b/g ad-hoc Kommunikation einzusetzen wohingegen die IEEE 802.11b/g Infrastrukturkommunikation mit den Sicherheitsmechanismen gemäß IEEE 802.11i [6] geschützt werden.

---

[5] Von etwa 10 MByte/s (LWL) bis nicht mehr als ein paar KByte/s (GPRS)
[6] IVS steht für ITS Vehicle Station und bezeichnet in sim$^{TD}$ die Fahrzeuge

## Zusammenfassung

Das verteilte IRS Management System bietet eine systemweite Verwaltung und Überwachung aller IRS im sim$^{TD}$ Versuchsgebiet. Die IRS verbindet in sim$^{TD}$ die beiden sehr unterschiedlichen Welten mit der verbindungslosen Fahrzeug Ad-Hoc Kommunikation auf der einen und der verbindungsorientierten Verkehrsmanagementsystemen mit allen ihren verschiedenen Protokollen auf der anderen Seite. Dabei bietet das IRS Management System eine effiziente und zuverlässige Umgebung für alle Testanwendungen. Das System garantiert einen 24/7 Betrieb und eine flüssige Bearbeitung von allen Aufgaben im Bezug auf die Anwendungen. Es ist ein selbstorganisierendes System, welches sich durch gezielte Anwendung von Strategien von einem Fehlerzustand automatisiert in den Normalbetrieb versetzen kann. Die aktuelle IRS Management System Architektur spiegelt alle wichtigen Aspekte wider, die für ein späteres System in der realen Welt notwendig sind. Hier sind vor allem Kommunikationsaspekte, das Management und die Wartung ebenso wie sinnvolle Verteilung und Organisation der Infrastrukturanwendungen zu nennen. Daher ist das vorgestellte IRS Management System nicht nur für die Durchführung eines Feldversuchs relevant, sondern wurde explizit für den Einsatz unter produktiven Bedingungen konzipiert und entwickelt.